\documentstyle[aps,prl,preprint]{revtex}
\begin{document}
\title{Mixed symmetry superconductivity in two-dimensional Fermi liquids}
\author{K.A. Musaelian, J. Betouras, A.V. Chubukov and R. Joynt}
\address{Department of Physics and Applied Superconductivity Center \\
University of Wisconsin-Madison\\
Madison, Wisconsin 53706\\}
\maketitle
\begin{abstract}
We consider a 2D isotropic Fermi
liquid with attraction in both $s$ and $d$ channels and
 examine the possibility of a superconducting state with mixed
$s$ and $d$ symmetry of the gap function.  We show that
both in the weak
coupling limit and at strong coupling, a mixed $s+id$ symmetry state
is realized in a certain range of
interaction. Phase transitions between
the mixed and the pure symmetry states are second order. We also show that
there is no stable mixed $s+d$ symmetry state at any coupling.
\end{abstract}

\newpage

\section{INTRODUCTION}
The question of the order parameter symmetry is one of the central issues of
high-temperature superconductivity.  There is a general consensus that the
superconducting gap is highly anisotropic, but whether the gap has a particular
symmetry under rotations is  still
a matter of debate. A number of experiments on YBCO-123 are roughly
consistent with the $d-$wave symmetry \cite{scala}
for which the most natural source is the exchange of magnetic
fluctuations~\cite{Pines}, but some experiments, e.g.,
photoemission studies on BSCCO-2212 \cite{marshall,ding} as well as $c$-axis
Josephson tunneling experiments on YBCO\cite{dynes},
are inconsistent with the pure d-wave, but more consistent with an $s + d$
state.  In principle, the presence
of orthorhombic distortion in, e.\ g.\ , BSCCO-2212 \cite{kirk} in itself
guarantees that an otherwise
$d$-wave superconducting gap will have an admixture of the $s$-wave
\cite{qpl}. However,
the superconducting state
may be a mixture of $s$ and $d$ components
even in the absence of an  orthorhombic distortion.

A superconducting state with a mixed $s +d$ symmetry of the gap
was first discussed in \cite{ruck}, and the $s+id$ state in \cite{kotliar}.
The mixed symmetry state at intermediate doping levels was also found in
variational Monte Carlo studies of the $t-J$ model\cite{ben}. An alternative
possibility of
symmetry mixing due to interplane coupling was proposed in Ref.\ \cite{lee}.
More exotic
mixed symmetry states have also been suggested \cite{laughlin,wiegmann}.

Very recent work has shown \cite{bec} that the extent of s-wave admixture
is a strong function of $t^{\prime}$,
 the second neighbor hopping, which varies a good
deal from one high-T$_c$ material to another.  This strongly suggests that
the question of s-d mixing should be looked at as a function of hole
doping, and that this must be done in each high-T$_c$ material separately.

The variational Monte Carlo calculations indicate that
at the doping levels which favor mixed symmetry states,
the ground state energy is roughly independent of the
relative phase $\theta$ of s- and d-wave components.  This relative phase is
of great importance, since only $\theta=0$ and $\theta=\pi$ states can have gap
nodes.
The experiments on combinations of Josephson junctions
on YBCO-123 \cite{vh} appear to rule out a relative phase
of $\pi/2$ if the s-wave amplitude is more than
about 10\% of the d-wave.  Recent photoemission
work as a function of hole doping \cite{marshall}, \cite{joe_bob}
indicates that $\theta = 0$ and that the relative amplitude of
s-wave and d-wave depends on temperature and hole doping level.

In view of both the experimental situation, which appears to indicate that
s-d mixing is possible in some systems and the theoretical situation where
the relative phase is not well-determined, additional understanding of the
physics of this phase is needed.
In this paper, we address the issue whether one can obtain the $s+d$ mixed
state
in calculations on a simple but fairly general model.
The answer we obtain is negative - we found
that for any coupling, the only possible mixed symmetry in this
model is $s + i d$.

We
consider a model of an isotropic two dimensional Fermi liquid
with attractive
interaction in both $s$ and $d$ channels. We assume that both interactions are
frequency-independent in a
frequency range
bounded by the cut-off frequency $\omega_c$, and are zero for $|\omega| >
\omega_c$.
Obviously, when
only one of the two interaction channels is present, the ground state
is described by the corresponding pure symmetry gap function. When both
interaction channels are present, their competition will lead to either one of
the two pure symmetry superconducting states, or a mixed state, where the gap
function contains both the $s$ and the $d$ harmonics.

In the next section we will
consider the weak coupling case, where one can use the BCS formalism. We will
show that the transition between $s$ and $d$ symmetries occurs via an
intermediate phase with mixed $s + id$ symmetry. The two phase transitions
between the pure and the mixed states are second order.
In Section III we consider the case of arbitrary coupling in the framework of
the Eliashberg theory. We will show that there always exists
a range of relative
strengths of the $s$ and $d$ interactions where an $s+id$ solution
exists. The analysis of the $s+d$ mixed state is more complicated. However, we
can show that at least in both the weak and the strong coupling limits the
$s+d$ mixed state does not occur. Our conclusions are summarized in Section IV.
As an aside, in the Appendix we also present few simple results for the
thermodynamics of a $d$-wave superconductor in the weak coupling limit, which
to the best of our knowledge have not been published anywhere else.
The main feature
is that the ratio of the superconducting gap to the transition
temperature, $2\Delta/T_c$, for
$d$-wave is 4.28, larger than  3.06 for the $s$-wave .

\section{WEAK COUPLING}
In this section we will consider the case when the coupling is weak in both
interaction channels. In this case the BCS theory is valid, and the
gap equation assumes the following form \cite{bcs:gap}
\begin{eqnarray}
\Delta(\vec{k})=-\sum_{\vec{k'}} V_{\vec{k}\vec{k'}}\frac{\Delta
(\vec{k'})}{2E_{\vec{k'}}}
\label{bcs_gap}
\end{eqnarray}
where
\begin{eqnarray}
E_{\vec{k}}=\sqrt{[\frac{{\hbar}^2 k^2}{2m}-\mu]^2 +|\Delta(\vec{k})|^2}
= \sqrt{\varepsilon_{\vec{k}}^2 +|\Delta(\vec{k})|^2}
\label{two}
\end{eqnarray}
and
\begin{equation}
V_{\vec{k}\vec{k'}} = V_s(\vec{k},\vec{k}')+V_d(\vec{k},\vec{k}')
\cos(2(\phi - \phi^{\prime}))
\end{equation}
is the interaction, which contains both $s$ and $d$ harmonics.
The angle $\phi$ is defined in our two-dimensional
model as $\phi=\tan^{-1}(k_y/k_x)$.
Consider now a trial mixed state with arbitrary phase difference between
the s-wave and the d-wave components of the order parameter:
\begin{eqnarray}
\Delta(\phi) = \Delta_s + e^{i\theta} \Delta_d \cos 2\phi ~,
\label{mixed_delta}
\end{eqnarray}
Separating the real and imaginary parts of this equation, and also $s$ and $d$
components, we obtain  a set of three independent
equations
\begin{eqnarray}
\Delta_{s} = - \sum_{\vec{k'}} {\frac{V_s (\Delta_s+\cos\theta\Delta_d
\cos 2\phi')}{2\sqrt{ {\epsilon_{\vec{k'}}}^2
+|\Delta(\vec{k'})|^2 } }}
\label{delta_s}\\
\Delta_d\cos\theta  = - \sum_{\vec{k'}} \frac{V_{d}\cos 2\phi'
(\Delta_{s}+\cos\theta \Delta_{d} \cos 2\phi')}
{2\sqrt{{\epsilon_{\vec{k'}}}
^2+ |\Delta(\vec{k'})|^2}}  \label{delta_d}  \\
\cos 2\phi\sin\theta = - \sum_{\vec{k'}} \frac{\sin\theta
\cos 2\phi'
( V_s+V_d \cos 2\phi \cos 2\phi')} {2\sqrt{{\epsilon_{\vec{k'}}}^2+
|\Delta(\vec{k'})|^2}}  \label{cossin}\\
\end{eqnarray}
where $V_s = V_s(k_F,k_F)$ and $V_d = V_d(k_F,k_F)$.
It is straightforward to see that if both $\Delta_s$ and $\Delta_d$ are finite,
the set (\ref{delta_s} - \ref{cossin})
can be simultaneously satisfied in only two cases
$\theta=0$ or $\theta = \pi/2$, leading to $s+d$ or $s+id$ respectively.
Thus the weak-coupling theory gives the same restricted set of possibilities
for the internal phase angle that Ginzburg-Landau theory offers~\cite{joe_bob}.
Below we consider these two cases separately.

\subsection{s+d state}

In this case $\theta = 0$, and equations (\ref{delta_s}) and (\ref{delta_d})
become :
\begin{eqnarray}
\Delta_{s} = - \int_{0}^{2\pi} \frac {d\phi'} {2\pi}
\int_{-\omega_{c}}^{\omega_{c}}
d\epsilon N(0)V_s \frac {\Delta_s+\Delta_d \cos 2\phi'} {4\sqrt{\epsilon^
2+\Delta_s^2+2\Delta_s\Delta_d\cos 2\phi +\Delta_d^2\cos^2 2\phi}}
\label{spd1}\\
\Delta_{d} = -
\int_{0}^{2\phi}\frac{d\phi'}{2\pi}\int_{-\omega_{c}}^{\omega_{c}}
d\epsilon N(0) V_d \frac{\cos 2\phi' (\Delta_s + \Delta_d
\cos 2\phi')}{4\sqrt{\epsilon^2 + \Delta_s^2 + 2\Delta_s\Delta_d\cos 2\phi
+\Delta_d^2\cos^2 2\phi}} ~.
\label{spd2}
\end{eqnarray}
Performing the integration over frequency and doing standard manipulations, we
obtain
\begin{equation}
(1 - \frac{2g_s}{g_d})\alpha = g_s f(\alpha)
\label{alpha}
\end{equation}
where $g_s = - V_sN(0)/4$, $g_d = - V_dN(0)/4$, $\alpha \equiv
\Delta_s/\Delta_d$, and the function $f(\alpha)$ is given by
\begin{equation}
f(\alpha) = \int_{0}^{\pi} \frac{dx}{2\pi} (2\alpha\cos x-1)(\cos x + \alpha)
\log (\alpha+\cos x)^2~.
\label{fofalpha}
\end{equation}
The graphical solution of Eq.(\ref{alpha}) is shown in Fig.\ \ref{fig_fal}.
It is easy to see that in the limit of $\alpha \to \infty$,
$f(\alpha) \sim \alpha$, while in the limit of $\alpha \to 0$, $f(\alpha) \sim
- \alpha/2$. If we start out with $g_s = 0$, then, naturally, the only solution
is $\alpha = 0$, i.e. pure $d$-wave. As $g_s$ increases, the slope of the
straight line on Fig.\ \ref{fig_fal} decreases, and at
\begin{equation}
g^{(1)}_s =  \frac{g_d}{2+g_d}~.
\label{d_to_s}
\end{equation}
the lines first cross at
$\alpha = \infty$.
If we increase $g_s$ even further, we find three solutions: at $\alpha =0,
{}~\alpha = \infty$, and at some finite $\alpha_m$, which corresponds to a
mixed
state.  As $g_s$ increases further, $\alpha_m$ decreases and becomes zero
at
\begin{equation}
g^{(2)}_s = \frac{g_d}{2-g_d/2}~.
\end{equation}
For larger $g_s$, there exists only one solution: $\alpha = \infty$ which
corresponds to a pure $s$-wave state.

We see therefore that $d-$wave solution exists at $0<g_s < g^{(2)}_s$ while the
$s-$wave solution exists at $g_s > g^{(1)}_s$. The key point is that
$g^{(2)}_s > g^{(1)}_s$, such that there is an
intermediate region $g^{(1)}_s <g_s < g^{(2)}_s$ where both pure solutions
exist
together with the $s + d$ solution. To verify which solutions are stable, we
computed the second derivatives of the energy and found
that the two pure solutions are stable in
the intermediate region while the $s + d$ state actually corresponds to a
maximum rather than a minimum of energy.
Clearly then, $s + d$ state is unstable; if it was the only
mixed state allowed, then the system would simply undergo
a first-order transition between
the two pure states with a region where hysteresis is possible
between $g^{(1)}_s$ and
$g^{(2)}_s$.

\subsection{s+id state}

In the case of $\theta=\pi/2$, we follow the same procedure. Now the coupled
gap equations have the following form
\begin{eqnarray}
\Delta_d = \int_{0}^{2\pi} \frac{d\phi'}{2\pi} \int_{-\omega_c}^{\omega_c}
d\epsilon g_d
\frac{\Delta_d\cos^2 2\phi'}{\sqrt{\epsilon^{2} + \Delta_s^2 +
 \Delta_d^2\cos^2{2\phi}}}
\label{sid_1} \\
\Delta_s = \int_{0}^{2\pi} \frac{d\phi'}{2\pi} \int_{-\omega_c}^{\omega_c}
d\epsilon g_s
\frac {\Delta_s}{\sqrt{\epsilon^{2} + \Delta_s^2 +
 \Delta_d^2\cos^2{2\phi}}}  ~.
\label{sid_2}
\end{eqnarray}
Integrating over frequency and doing standard manipulations we obtain that the
mixed state exists if
\begin{equation}
1 - \frac{2g_s}{g_d} = g_s\left(
\frac{1}{2}+\alpha^2-\alpha\sqrt{1+\alpha^2}\right).
\label{alpha:sid}
\end{equation}
As before, $\alpha = \Delta_s/\Delta_d$.

As we start out with $g_s = 0$, Eq.\ (\ref{alpha:sid}) has no solution and the
gap
has pure $d$ symmetry $(\alpha \equiv 0$). However, contrary to the previous
case, here a  solution of (\ref{alpha:sid}) first appears
at
\begin{equation}
g^{(1i)}_s = \frac{g_d}{2+g_d/2}~,
\end{equation}
for the same $\alpha = 0$. As we increase $g_s$ from $g^{(1i)}_s$,
$\alpha$ and therefore $\Delta_s$ increases continuously, satisfying
\begin{equation}
\alpha = \frac{g_s/2+2g_s/g_d-1}{\sqrt{2g_s\left(1-
2g_s/g_d\right)}}~,
\end{equation}
and becomes infinite at $g^{(2i)}_s = g_d/2$. Clearly,
in this situation, we have
a second order phase transition from pure $d$ to a mixed
$s+id$ symmetry state at $g_s = g^{(1i)}_s$, and a second order transition
from a mixed state to pure $s$ state at $g_s = g^{(2i)}_s > g^{(1i)}_s$.
In other words, the two pure states, which are stable with respect to $s + d$
mixture, are in fact unstable (for corresponding $g_s$)
with respect to $s + id$ mixture, and in between $g^{(1i)}_s$ and $g^{(2i)}_s$,
the $s + id$ state is the equilibrium state of a system (see
Fig.\ref{fig_phase}).

\section{STRONG COUPLING}

In order to be certain that our results are not an artifact of the
weak-coupling
approximation,
we perform the calculations in the strong coupling regime.  We follow the
Eliashberg formalism \cite{eliashberg,scalapino} at zero temperature. We assume
that the frequency cutoff $\omega_c \ll \epsilon_F$,
so that vertex corrections can be
neglected according to Migdal theorem \cite{migdal}.

In the Eliashberg approach, one preserves the frequency dependence of the gap,
and substitutes the full quasiparticle Green function in the gap equation.
In explicit form, the equations are
\begin{eqnarray}
\tilde{\Delta}(\vec{k},\omega)= - \int \frac{d\omega'}{2\pi} \sum_{\vec{k'}}
V_{\vec{k}\vec{k'}}\frac{\tilde{\Delta}(\vec{k'},\omega')}
{\Omega^2(\vec{k'},\omega') +{\tilde{\Delta}}^2(\vec{k'},\omega') +
 \xi^2(\vec{k'})}  \nonumber\\
\Omega(\vec{k},\omega)= \omega - \int \frac{d\omega'}{2\pi}\sum_{\vec{k'}}
V_{\vec{k}\vec{k'}}
\frac{\Omega(\vec{k'},\omega')} {\Omega^2(\vec{k'},\omega') +
\tilde{\Delta}^2(\vec{k'},\omega') + \xi^2(\vec{k'})} ~,
\label{eliash}
\end{eqnarray}
where  $\xi(\vec{q}) $ is the renormalized single particle energy, $\xi(q) =
v_F (q - p_F),~
\tilde{\Delta}(\vec{q},\omega)$ is the anomalous  part of the self-energy, and
$\Omega (\vec{q},\omega)$ is the antisymmetric in $\omega$ part of the normal
self-energy. As $\omega_c \ll \epsilon_F$, the integration over
$|k^{\prime}|$ is confined to a region near the Fermi surface and can be
substituted by the integration over $\xi$, as in BCS theory.
This integration is straightforward,
and yields
\begin{eqnarray}
\tilde{\Delta}(\omega)&=&\int_{\omega-\omega_{0}/2}^{\omega+\omega_{0}/2}
d\omega' \int_{0}^{2\pi} \frac{d\phi}{2\pi} \frac{(g_{s}+g_{d} \cos2\phi)
\tilde{\Delta}(\omega',\phi)}
 {\sqrt{|\Omega(\omega',\phi)|^2+
|\tilde{\Delta}(\omega',\phi)|^2 }}  \nonumber\\
\Omega(\omega,\phi)&=& \omega+\int_{\omega-\omega_{0}/2}^{\omega+\omega_{0}/2}
d\omega' \int_{0}^{2\pi} \frac{d\phi}{2\pi} \frac{(g_{s}+g_{d} \cos2\phi)
\Omega
(\omega',\phi)}{\sqrt{|\Omega(\omega',\phi)|^2+|\tilde{\Delta}
(\omega',\phi)|^2 }}
\label{eliash2}
\end{eqnarray}
We now again consider the two mixed states separately.

\subsection{$s+id$ state}

Consider first a mixed $s+id$ state.  For this symmetry, the
angular decomposition of the self-energy functions yields
\begin{eqnarray}
\tilde{\Delta}(\omega,\phi) = \tilde \Delta_s(\omega) +
 i \tilde \Delta_d(\omega) \cos(2\phi)  \\
\Omega(\omega,\phi) = \Omega_s(\omega) + i \Omega_d(\omega) \cos(2\phi)~.
\end{eqnarray}
Accordingly, Eqs.(\ref{eliash2}) can be decomposed into four equations
\begin{eqnarray}
\tilde \Delta_d(\omega)&=&\int_{\omega-\omega_0/2}^{\omega+\omega_0/2}
d\omega' \int_0^{2\pi} \frac{d\phi}{2\pi}\frac{g_d\tilde \Delta_d
(\omega')\cos^2 2\phi}{\sqrt{\Omega_s^2(\omega')+\Omega_d^2(\omega') +
\tilde\Delta_s^2(\omega') + \tilde\Delta_d^2(\omega')\cos^2 2\phi}}
\label{delta_d2}\\
\tilde \Delta_s(\omega)&=&\int_{\omega-\omega_0/2}^{\omega+\omega_0/2}
d\omega' \int_0^{2\pi} \frac{d\phi}{2\pi}\frac{g_s\tilde \Delta^s
(\omega')}{\sqrt{\Omega_s^2(\omega')+\Omega_d^2(\omega') +
\tilde\Delta_s^2(\omega') + \tilde\Delta_d^2(\omega')\cos^2 2\phi}}
\label{delta_s2}\\
\Omega_d(\omega)&=& \int_{\omega-\omega_{0}/2}^{\omega+\omega_{0}/2}
 d\omega' \int_{0}^{2\pi} \frac{d\phi}{2\pi} \frac{g_d\Omega_d
(\omega')\cos^2 2\phi}{\sqrt{\Omega_s^2(\omega')+\Omega_d^2(\omega') +
\tilde\Delta_s^2(\omega') + \tilde\Delta_d^2(\omega')\cos^2 2\phi}}
\label{omega_d}\\
\Omega_s(\omega)&=&\omega +\int_{\omega-\omega_{0}/2}^{\omega+\omega_{0}/2}
 d\omega' \int_{0}^{2\pi} \frac{d\phi}{2\pi} \frac{g_s\Omega_s
(\omega')}{\sqrt{\Omega_s^2(\omega')+\Omega_d^2(\omega') +
\tilde\Delta_s^2(\omega') + \tilde\Delta_d^2(\omega')\cos^2 2\phi}}
\label{omega_s}
\end{eqnarray}
Equation (\ref{omega_d}) is homogeneous in $\Omega_d$. For
weak coupling its only solution was $\Omega_d \equiv 0$. In principle, at
strong coupling, there is a chance that above some threshold,
there exists a solution with a non-zero $\Omega_d$. We will not consider
this rather exotic possibility  and assume that the stable solution of
Eq.(\ref{omega_d}) corresponds to $\Omega_d = 0$ for all couplings.

We now follow the same approach as at weak coupling and
look for the transition points between pure $s$ and mixed $s+id$, and
pure $d$ and mixed $s+id$ states. In the former case,
 we linearize the above set of
equations around $\tilde\Delta_d =0$ and obtain
\begin{eqnarray}
\tilde\Delta_d(\omega) = \int_{\omega-\omega_0/2}^{\omega+\omega_0/2}d\omega'
\int_0^{2\pi} \frac{d\phi}{2\pi}\frac{g_d\tilde\Delta_d(\omega')\cos^2
2\phi}{\sqrt{\Omega_s^2(\omega')+\tilde\Delta_s^2(\omega')}}\nonumber\\
\tilde\Delta_s(\omega) = \int_{\omega-\omega_0/2}^{\omega+\omega_0/2}d\omega'
\int_0^{2\pi} \frac{d\phi}{2\pi}\frac{g_s\tilde\Delta_s(\omega')}
{\sqrt{\Omega_s^2(\omega')+\tilde\Delta_s^2(\omega')}}
\label{s_to_m}
\end{eqnarray}
Eqs.(\ref{s_to_m}) are obviously satisfied when $g^{(2i)}_s = g_d/2$,
 the same as for weak coupling.

Now consider the transition from the $d$-wave state into the mixed state.
Linearizing Eqs.(\ref{delta_d2})-(\ref{omega_s}) with respect to
$\tilde\Delta_s$ we get
\begin{eqnarray}
\tilde\Delta_d(\omega) = \int_{\omega-\omega_0/2}^{\omega+\omega_0/2}d\omega'
\int_0^{2\pi}
\frac{d\phi}{2\pi}\frac{g_d\tilde\Delta_d(\omega')}
{2\sqrt{\Omega_s^2(\omega')+\tilde\Delta_d^2(\omega')\cos^2 2\phi}}\nonumber\\
+ \int_{\omega-\omega_0/2}^{\omega+\omega_0/2}d\omega'
\int_0^{2\pi} \frac{d\phi}{2\pi}\frac{g_d\tilde\Delta_d(\omega')\cos 4\phi}
{2\sqrt{\Omega_s^2(\omega')+\tilde\Delta_d^2(\omega')\ cos^2 2\phi}}
\label{d_to_md}\\
\tilde\Delta_s(\omega) = \int_{\omega-\omega_0/2}^{\omega+\omega_0/2}d\omega'
\int_0^{2\pi}\frac{d\phi}{2\pi}\frac{g_s\tilde\Delta_s(\omega')}
{\sqrt{\Omega_s^2(\omega')+\tilde\Delta_d^2(\omega')\cos^2 2\phi}}
\label{d_to_ms}
\end{eqnarray}

We first observe that if the second term in
Eq. (\ref{d_to_md}) were absent, the $d-$ wave solution would become unstable
at
$g^{(1i)}_s = g_d/2 = g^{(2i)}_s$. We now show that this second term yields a
negative correction to the first term independent of what ${\tilde \Delta}_d
(\omega^{\prime})$ and $\Omega_s (\omega^{\prime})$ are. Indeed, let's perform
angular integration first. For the first term in (\ref{d_to_md}), the integrand
(apart from $g_d\tilde\Delta_d(\omega')$) is positive, and the integration
yields a positive result. For the second term, we have to evaluate
\begin{equation}
I = \int_0^\pi \frac{\cos \phi d\phi}{\sqrt{a+b\cos\phi}}~,
\end{equation}
where $a > b > 0$. Doing simple manipulations, we obtain
\begin{equation}
I = \int_0^{\pi/2} d\phi\left(\frac{\cos\phi}{\sqrt{a+b\cos\phi}}-
\frac{\cos\phi}{\sqrt{a-b\cos\phi}}\right)~.
\end{equation}
Since $\cos\phi$ is positive when $0<\phi<\pi/2$, then
 $I$ is negative.
This simple argument shows that the second term in
(\ref{d_to_md}) effectively reduces the value of $g_d$ to $g^{eff}_d < g_d$.
 Clearly then,
the critical value  $g^{(1i)}_s = g^{eff}_d/2$ for the transition between the
pure $d$ state and the $ s + id$ mixed state will be {\it smaller} than
$g_d/2 \equiv g^{(2i)}$.

The above consideration shows that for arbitrary interaction strength,
exists a range of $g_s/g_d$ where neither of the pure states is
stable. We further expanded near each of the transition points
up to qubic terms in either $\Delta_s$ or $\Delta_d$,
and indeed found a nonzero
solution for this intermediate range of couplings which implies that
in the intermediate region, the gap has an $s + id$ symmetry. It turns out
therefore that the phase diagram at weak and strong couplings is
essentially the same.  It is nevertheless interesting that the strong
coupling corrections tend make the mixing of s-wave into
a predominantly d-wave more favorable.

\subsection{$s + d$ state}

Now we study whether it is possible to
obtain the $s+d$ mixed symmetry state in equilibrium.
This case is significantly more
complicated, because unlike the $s+id$ case, the square of the
gap function now contains a term which is linear in both  $s$
and $d$ components.

 However, we will show that at least in the limit of very strong coupling,
there is no region of $s+d$ symmetry. Indeed,
consider the transition between the pure
$s$ and the mixed $s+d$ states. Eliashberg equations linearized in
$\tilde\Delta_d$ and $\Omega_d$  read
\begin{eqnarray}
\tilde\Delta_s(\omega) = \int_{\omega-\omega_0/2}^{\omega+\omega_0/2}d\omega'
\int_0^{2\pi}
\frac{d\phi}{2\pi}\frac{g_d\tilde\Delta_s(\omega')}
{\sqrt{\Omega_s^2(\omega')+\tilde\Delta_d^2(\omega')}}
\label{s_to_ms}\\
\Omega_s(\omega) = \omega +
\int_{\omega-\omega_0/2}^{\omega+\omega_0/2}d\omega'
\int_0^{2\pi}
\frac{d\phi}{2\pi}\frac{g_s\Omega_s(\omega')}{\sqrt{\Omega_s^2(\omega'))+
\tilde\Delta_d^2(\omega')}}\label{s_to_mos}\\
\tilde\Delta_d(\omega) = \int_{\omega-\omega_0/2}^{\omega+\omega_0/2}d\omega'
\int_0^{2\pi} \frac{d\phi}{2\pi}\frac{g_d\tilde\Delta_d(\omega')\cos^2 2\phi}
{\sqrt{\Omega_s^2(\omega')+\tilde\Delta_d^2(\omega')\cos^2 2\phi}}\nonumber\\
-
\frac{g_d\tilde\Delta_d(\omega')(\tilde\Delta_d(\omega')\tilde\Delta_s(\omega')
+
\Omega_d(\omega')\Omega_s(\omega'))\cos^2 2\phi}{2(\Omega_s^2(\omega')+
\tilde\Delta_s^2(\omega'))^{3/2}}\label{s_to_md}\\
\Omega_d(\omega) = \int_{\omega-\omega_0/2}^{\omega+\omega_0/2}d\omega'
\int_0^{2\pi}
\frac{d\phi}{2\pi}\frac{g_d\Omega_d(\omega')}
{\sqrt{\Omega_s^2(\omega)+\tilde\Delta_d^2(\omega)\cos^2 2\phi}}\nonumber\\
-
\frac{g_d\tilde\Delta_d(\omega')(\tilde\Delta_d(\omega')\tilde\Delta_s(\omega')
+
\Omega_d(\omega')\Omega_s(\omega'))\cos^2 2\phi}{2(\Omega_s^2(\omega')+
\tilde\Delta_s^2(\omega'))^{3/2}}
\label{s_to_mod}
\end{eqnarray}
We now show that for $g_s \gg 1$ and
$\omega \leq \omega_0$, $\tilde\Delta_s(\omega) \gg \Omega_s(\omega)$.
Indeed, suppose that this is true. Then it is easy to see that in
the region of frequencies we are interested in,
$\tilde\Delta_s(\omega)$ is frequency independent and
equal to $g_s\omega_0$, while $\Omega_s (\omega)$ is linear in $\omega$.
Letting $\Omega_s (\omega) = \lambda\omega$, we can rewrite
the inequality  as $g_s \gg \lambda$. Then, equation (\ref{s_to_mos})
becomes
\begin{equation}
\lambda = 1 +
\frac{g_s\lambda}{\omega}\int_{\omega-\omega_0/2}^{\omega+\omega_0/2}
\frac{\omega'd\omega'}{\sqrt{\lambda^2{\omega'}^2+\tilde\Delta_s^2}}~.
\label{lambda1}
\end{equation}
Solving (\ref{lambda1}) we obtain for $g_s \gg 1$
\begin{equation}
\lambda = 2 g_s^{2/3} \ll g_s
\label{lambda2}
\end{equation}
thus justifying the assumption that
$\tilde\Delta(\omega) \gg \Omega_s(\omega)$.
Furthermore, in this limit, Eq.(\ref{s_to_md}) reduces to
\begin{equation}
1 = \frac{g_d}{2}\int_{-\omega_0/2}^{\omega_0/2}\frac{\lambda^2\omega^2
d\omega}{(\tilde\Delta_s^2 + \lambda^2\omega^2)^{3/2}}
\label{g_d}
\end{equation}
Using (\ref{lambda2}) we finally obtain that
\begin{equation}
g_d = 6 g_s^{5/3} \gg g_s ~.
\label{g_d2}
\end{equation}
We see therefore that in the strong coupling limit, the hypothetical transition
between a mixed $s+d$ and a pure $s$ states occurs at $g_s = g^{(2)}_s
 \ll g_d$.
However, for this ratio of couplings, both the pure $s$-wave and
the mixed state clearly must be unstable with respect
to the pure $d$-wave state.
A similar analysis shows that the transition between the pure $s$ and the mixed
state occurs at $g_s = g^{(1)}_s \sim g_d$. As a result, we again have
$g^{(1)}_s > g^{(2)}_s$, which, just as in the weak coupling case,
implies that there is no region of mixed $s+d$ symmetry.

\section{SUMMARY}

We have studied in this paper a two-dimensional
isotropic Fermi liquid with attractive interaction in both
$s$ and $d$ channels. We considered  the weak coupling limit and also applied
the Eliashberg formalism at strong coupling.

The phase diagram of
the superconductor turns out to be the same at weak and strong coupling.
It displays a region with a mixed $s+id$ symmetry gap when the
coupling strengths in the two channels are of the same order of magnitude. The
phase transitions between the mixed state and the pure $s$ and $d$ states are
second order.

We have shown that in the weak and in the strong coupling limits a mixed
$s+d$ state does not occur. Intuitively, this can be interpreted as the
propensity of the system to choose the state in which the
amplitude of the gap function has the largest value.
This is also in agreement with the
Ginsburg-Landau considerations \cite{lee,joe_bob}, which suggest that in the
absence of orthorhombic distortion, the $s+id$ state
has lower energy than the $s+d$ state.

Indeed, the model we considered is oversimplified: the
 two-dimensional isotropic Fermi liquid captures some
 of features of the high-T$c$
materials,  however, the isotropic system is never close to a magnetic or
a metal-insulator transition.
Our analysis of the $s+d$ versus $s+id$ question therefore
does not include such effects.
In this sense, what we have shown here is
 that if the $s+d$ state is realized in real
materials, some nontrivial physical effect of the proximity to these phase
transitions is likely to be the cause.

\section{ACKNOWLEDGEMENTS}

This work was supported by the
NSF Materials Research Group Program under
Grant No. DMR-9214707 and by Grant No. DMR-9214739 .

\section{APPENDIX}

In this Appendix,
 we present weak coupling calculations for the thermodynamics of a pure
$d$-wave superconductor.

The gap equation at zero temperature is
\begin{equation}
1 ~=~
g_d\int_{-\omega_0/2}^{\omega_0/2}d\epsilon\int_0^{2\pi}\frac{d\phi}{2\pi}
\frac{\cos^2 2\phi}{\sqrt{\epsilon^2+\Delta_d^2\cos^2 2\phi}}~.
\end{equation}
After simple manipulations we arrive at
\begin{equation}
\Delta_d = \frac{2\omega_0}{\sqrt{e}}\exp\left(-\frac{1}{g_d}\right)~.
\end{equation}
At the same time, the transition temperature is given by ~\cite{bcs:gap}
\begin{equation}
T_c = \frac{\gamma\omega_0}{\pi}\exp\left(-\frac{1}{g_d}\right)~,
\end{equation}
where $\log\gamma = C \approx 0.577$ is Euler's constant. Then
\begin{equation}
\frac{2\Delta_d}{T_c} = \frac{4\pi}{\sqrt{e}\gamma} \approx 4.28~.
\end{equation}
Note, this ratio for $s$-wave is 3.06.

Finally, starting with the gap equation at finite temperature
\begin{equation}
1 ~=~ g_d\int_{-\omega_0/2}^{\omega_0/2} d\epsilon~
\int_0^{2\pi}\frac{d\phi}{2\pi}
\frac{\cos^2 2\phi}{\sqrt{\epsilon^2+\Delta_d^2}} = 1~,
\end{equation}
and performing standard expansions at $T \ll T_c$ and $T_c-T \ll T_c$
{}~\cite{bcs:gap} we find the following expressions for the temperature
dependence of the superconducting gap
\begin{eqnarray}
\Delta_d(T) = \Delta_d(0)\left[1-0.37\left(\frac{T}{T_c}\right)^3\right]
{}~~~~& T \ll T_c\nonumber\\
\Delta_d(T) = 1.65\Delta_d(0)\sqrt{1-T/T_c}~~~~& T_c - T \ll T_c~.
\end{eqnarray}

\begin{figure}
\caption{Graphical solution of Eq.\ (12) is given by the
intersection of the curve $f(\alpha)$ and the straight line.}
\label{fig_fal}
\end{figure}

\begin{figure}
\caption{The location of the phase boundaries for the trial $s+d$ state.
The   critical point of the transition from $s+d$ to $s$ occurs inside the
d-phase and vice versa, meaning that there is no region of $s+d$ mixed phase.}
\label{hyster}
\end{figure}

\begin{figure}
\caption{The phase diagram of the superconductor at zero temperature depending
on the ratio of coupling strengths in the two channels.}
\label{fig_phase}
\end{figure}

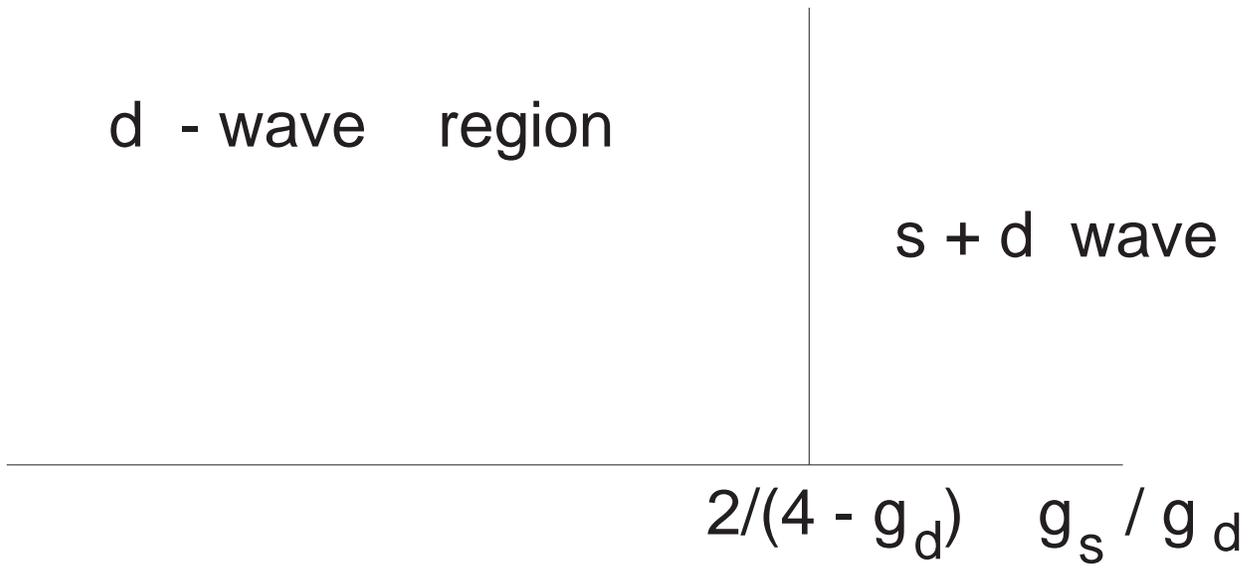
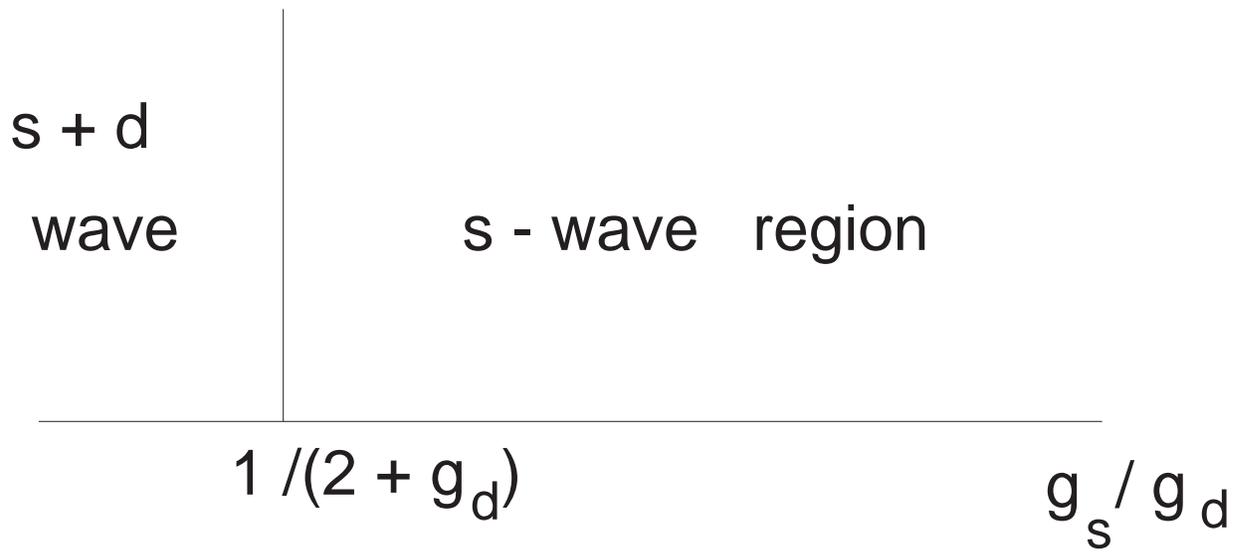
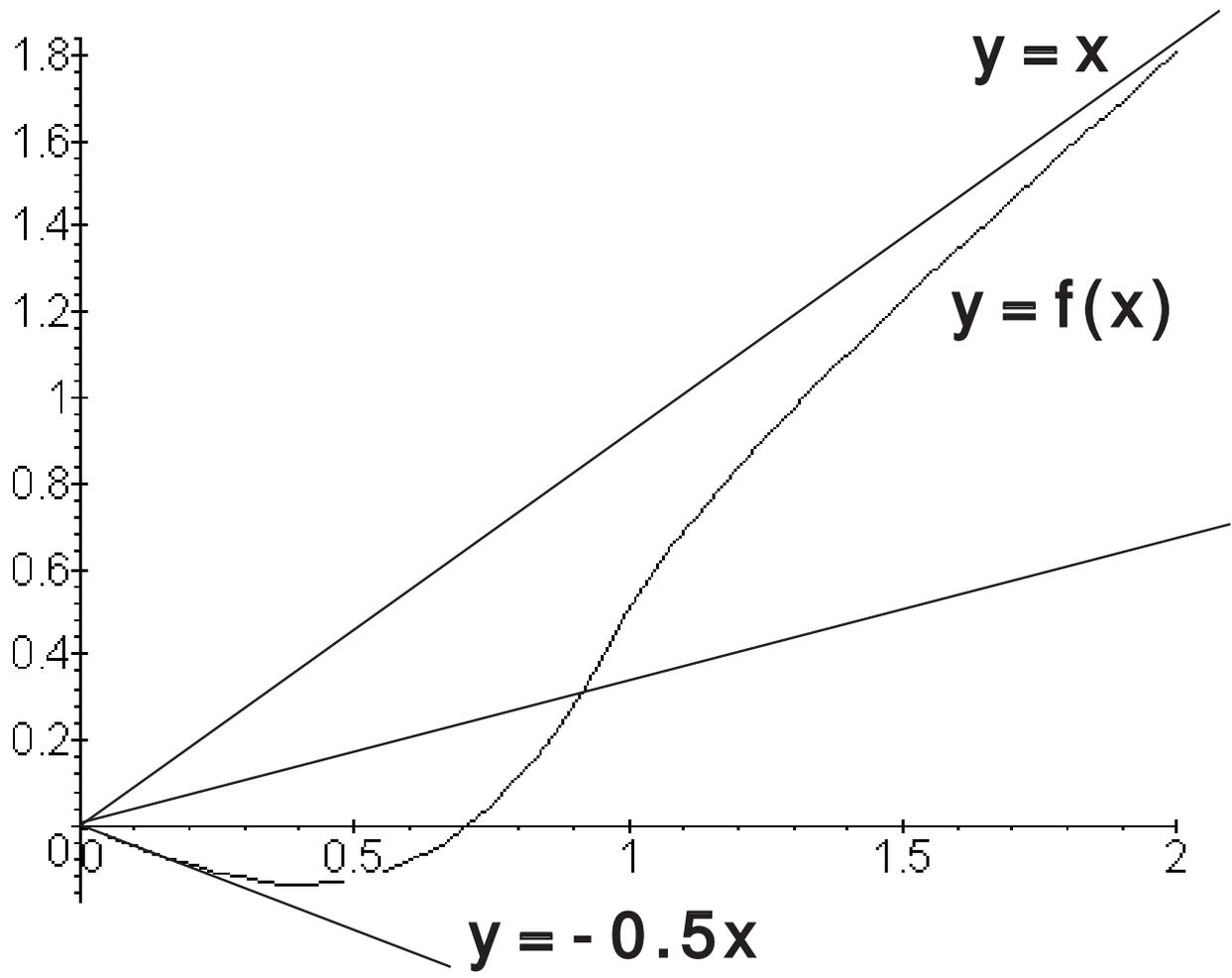
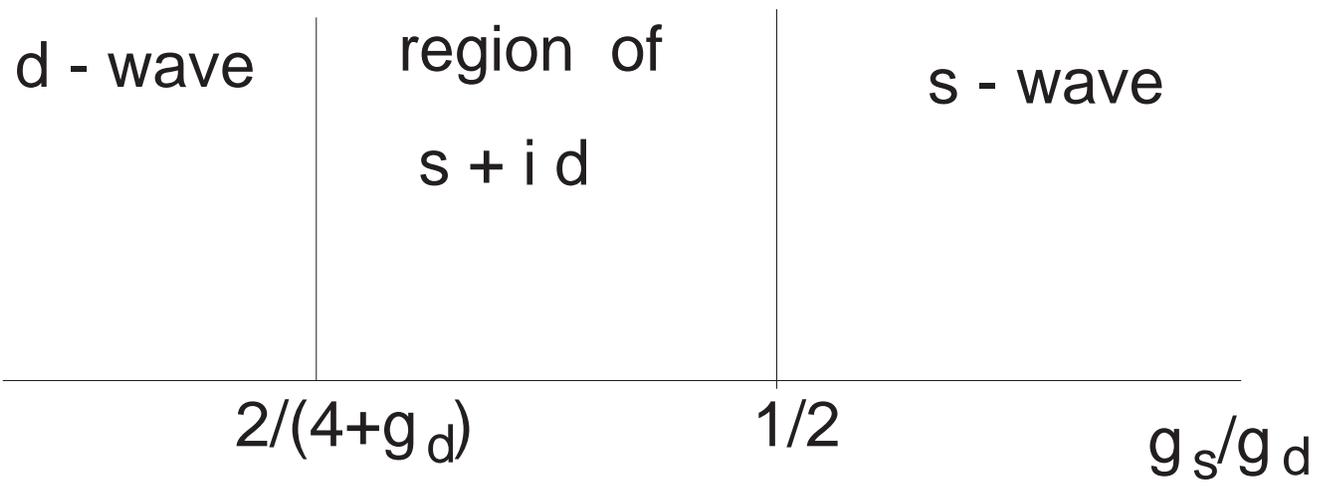
\end{document}